\documentclass[pra,twoside,twocolumn,showpacs,showkeys,floatfix]{revtex4}
\usepackage{times,graphicx,bbm,amsmath}
\def\smg{{\hbox{\small\sc s}}}

\def\sm{{\hbox{\scriptsize\sc s}}}
\def\pm{{\hbox{\scriptsize\sc p}}}
\begin{document}
\title{Information/disturbance trade-off in continuous variable Gaussian systems}
\author{Marco G. Genoni and Matteo G. A.  Paris}
\email{matteo.paris@fisica.unimi.it}
\affiliation{Dipartimento di Fisica dell'Universit\`a di Milano, Italia.}
\date{\today}
\begin{abstract}
We address the information/disturbance trade-off for state-measurements 
on continuous variable Gaussian systems and suggest minimal schemes 
for implementations. In our schemes, the symbols 
from a given alphabet are encoded in a set of Gaussian signals which 
are coupled to a probe excited in a known state. After the interaction the probe
is measured, in order to infer the transmitted state, while the
conditional state of the signal is left for the subsequent user.
The schemes are minimal, {\em i.e.} involve a single additional probe,
and allow for the nondemolitive transmission of a continuous real
alphabet over a quantum channel. The trade-off between information
gain and state disturbance is quantified by fidelities and, after
optimization with respect to the measurement, analyzed in terms of
the energy carried by the signal and the probe.
We found that transmission fidelity only depends on the energy
of the signal and the probe, whereas estimation fidelity also depends
on the alphabet size and the measurement gain. Increasing the probe
energy does not necessarily lead to a better trade-off, the most
relevant parameter being the ratio between the alphabet size and
the signal width, which in turn determine the allocation of the
signal energy.
\end{abstract}
\maketitle
\section{Introduction}\label{s:intro}
In a multiuser transmission line each user should decode the transmitted 
symbol and leave the carrier for the subsequent user. Therefore, some 
device is needed that, at each use of the a channel, permits the retrieval 
of information without the destruction of the carrier. 
In a quantum channel symbols are encoded in states of a physical system 
and therefore the ultimate bounds on the channel performances are posed by 
quantum mechanics. Indeed, any measurement aimed to extract information 
on a quantum state alters the state itself, {\em i.e.} produces a
disturbance \cite{hh0}. Quantum information, in fact, cannot be perfectly 
copied, neither locally \cite{nocl} nor at distance \cite{telecl}. 
Overall, there is an information/disturbance trade-off which unavoidably
limits the accuracy independently on the coding scheme \cite{KB} 
\par
Several approaches have been proposed to face this 
problem, either based on measuring (destructively) and partially 
recreating the signal \cite{gra,hau}, sharing entanglement
over large distances  \cite{briegel,child1,child2,kok} or pairing
coding \cite{shi}. The above schemes are referred to as quantum
repeaters. 
\par
In this paper we address devices which, besides extracting information, 
preserves, at least in part, the entire quantum state of the 
signal, {\em i.e.} the statistics of {\em all} possible 
observables. Our device thus conveys characteristics of both 
quantum nondemolition (QND) measurements of a given observable, 
and classical repeaters, whose goal is to preserve the global
information carried by a signal. Since the main feature of our 
scheme is the tunability of the information-disturbance 
trade-off \cite{Fid}, without any specific focus on the measurement,  
we do not refer to them as QND schemes, whose goal is limited to preserve 
the statistics of a specific observable. 
\par
The trade-off between information gain and quantum state
disturbance can be quantified in different ways \cite{hh1}, here
we use fidelities, which may be
defined as follows.  Suppose one wants to transmit the symbol $a$,
chosen from the alphabet ${\mathbbm A}$ according to the
probability density $p(a)$. To this aim a quantum system
is prepared in the pure state $|\psi_a\rangle$, chosen from a given set,
and then transmitted along a given channel. In order to share the
information among several users one needs a device which couples the signal to
one or more probe systems in order to produce two outputs. One of the
two outputs is sent to a user, who measures a predetermined
observable to infer the transmitted state, whereas the
(conditional) state of the second output is left to the subsequent
user and thus should contain an approximate copy of the input
signal. If the outcome $b$ is observed after the device, then
the estimated signal state is given by $|\phi_b\rangle$ (a {\em
natural} inference rule being $b\rightarrow |\phi_b\rangle$ with
$|\phi_b\rangle$ given by the set of eigenstates of the measured
observable), whereas the conditional state $|\varphi_b\rangle$ is
left for the subsequent user. The amount of disturbance is
quantified by evaluating the overlap of the conditional state
$|\varphi_b\rangle$ to the initial one $|\psi_a\rangle$, whereas
the amount of information extracted by the measurement corresponds
to the overlap of the inferred state $|\phi_b\rangle$ to the
initial one.  The corresponding fidelities, for a given input
signal $|\psi_a\rangle$, are given by
\begin{eqnarray}
F_a &=& \int_{\mathbbm B} db\: q(b)\:
\left|\langle\varphi_b|\psi_a\rangle\right|^2
\label{F_psi}
\\ G_a &=& \int_{\mathbbm B} db\: q(b)\:
\left|\langle\phi_b|\psi_a\rangle\right|^2
\label{G_psi}\;,
\end{eqnarray}
where we have already performed the average over the distribution
$q(b)$ of the outcomes.  The alphabet ${\mathbbm B}$ of the
output symbols ({\em i.e.} the spectrum of the measured
observable) is not necessarily equal to the input one, though
this choice is an optimized one \cite{rmp}.  The relevant quantities
to assess the performances of the scheme are then given by the
average fidelities
\begin{align}
F &=  \int_{\mathbbm A} \int_{\mathbbm B} da\, db \:
p(a)\:q(b)\: \left|\langle\varphi_b|\psi_a\rangle\right|^2
\\
G &=  \int_{\mathbbm A} \int_{\mathbbm B} da\, db \:
p(a)\:q(b)\: \left|\langle\phi_b|\psi_a\rangle\right|^2
\end{align}
which are obtained by averaging $F_a$ and $G_a$ over the possible
input states, {\em i.e.} over the alphabet ${\mathbbm A}$ of
transmittable symbols. $F$ will be referred to as the
\emph{transmission fidelity} and $G$ as the \emph{estimation
fidelity}. Of course we have $0\leq G \leq 1$ and $0\leq F \leq 1$
with $F=1$ corresponding to zero disturbance and $G=1$ to complete
information.
\par
Our device is local and its action is independent on the presence
of losses along the transmission line. The presence of losses before 
the device degrades the signals and, as a matter of fact, is equivalent 
to consider a set of mixed states at the input. The performance of 
such a device can be obtained from the analysis of
the pure state case by averaging the fidelity over the input probability. 
The result is an overall degradation of performances.
Since our focus is on exploring the ultimate bounds imposed by
quantum mechanics, we are not going to take into account mixing
at the input. In the following, we consider scheme suitable 
to transmit a continuous alphabet ${\mathbbm A}$ , whose symbols are
encoded in a set of Gaussian pure states of a continuous variable
(CV) infinite dimensional system.
\par
Let us first consider the extreme case: if nothing is done, the
signal is preserved $\forall a$ and thus $F=1$. However, at the same time,
our estimation has to be random, and thus $G\rightarrow 0$ since
we are dealing with an infinite-dimensional systems.  This
corresponds to a {\em blind} regeneration scheme \cite{sze}, which
re-prepares any quantum state received at the input, without
gaining any information on it. The opposite case is when the
maximum information is gained on the signal, {\em i.e.} when the
optimal estimation strategy for the parameter of interest is
adopted \cite{hel}. In this case $G\neq 0$, but then
the signal after this operation cannot provide any more information
on the initial state.  Our aim is to study intermediate cases,
{\em i.e.} quantum measurements providing only partial information
while partially preserving the quantum state of the signal for
subsequent users.  These kind of schemes, which correspond to
feasible quantum measurements, may be also viewed as universal 
quantum nondemolition measurements \cite{bra} ({\em i.e} not build for a 
specific observable), which have been widely investigated for
CV systems, and recently received attention also
for qubits \cite{dec04}.
\par
For discrete variable, the trade-off between information gain and
state disturbance  has been explicitly evaluated \cite{KB}, as
well as the bound that fidelities should satisfy according to
quantum mechanics. In turn, optimal schemes 
finite-dimensional systems (qudits), {\em i.e.} devices whose
fidelity balance saturates the bound have been suggested
\cite{Fid,filip} (in Ref. \cite{Fid} those schemes have been referred to 
as optimal quantum repeaters).
\par
As a matter of fact the fidelity bound for finite-dimensional
systems cannot be straightforwardly extended to infinite dimension,
and no analogue bound has been derived for CV systems, except for
the case of coherent states in phase-insensitive device \cite{ulr}
and non Gaussian protocol \cite{mst}.
Therefore,
in order to gain insight on the fidelity balance for CV systems
and to clarify the role of energy allocation, in this paper we suggest a class of
{\em minimal} schemes which involve
a single additional probe, and evaluate their performances
as a function of the channel (signal and probe) energy, which, in
turn, depends on the {\em width} of the signal and the probe
wave-packets as well as the {\em size} of the alphabet. Indeed
energy constraints are the main focus in infinite-dimensional
systems, and may serve to define optimality \cite{nap}
\par
The paper is structured as follows. In Section \ref{s:sz} we
describe our schemes and evaluate the probability of the
outcomes as well as the corresponding conditional states, whereas
in Section \ref{s:fid} we evaluate fidelities and analyze the
information-disturbance trade-off in terms of the signal and
the probe energy for different configurations. 
In Section \ref{s:imp} we discuss the optical implementation
of our schemes, whereas Section \ref{s:out}
closes the paper with some concluding remarks.
\section{Continuous variable indirect measurements}\label{s:sz}
In this section we suggest a measurement scheme suitable to infer
the information carried by a class of Gaussian CV states without
destroying the signals themselves. The setup is the
generalization to infinite dimension of the optimal schemes
suggested in \cite{Fid} for qudits. The
setup is minimal because it involves a single additional probe
system.
\subsection{Input signals}
We consider the transmission of a real alphabet ${\mathbbm A}
\equiv {\mathbbm R}$ with symbols $a$ encoded in the set of
Gaussian signals
\begin{eqnarray}
|\psi_{a\tau}\rangle_\sm = \int dx\: g_{a,\tau}(x)|x\rangle_\sm
\label{signal}
\end{eqnarray}
where  $|x\rangle$ denotes the standard CV basis, say position
eigenstates, with $\langle x'|x\rangle=\delta(x-x')$ and
$$|g_{a,\tau}(x)|^2 =
\frac{1}{\sqrt{2\pi}\tau} \exp\left[-\frac{(x-a)^2}{2\tau^2}
\right]\:.$$
The label $\smg$ indicates signal quantities throughout the paper.
We assume, without loss of generality, $g_{a,\tau}(x)$ as real,
{\em i.e.} that the signal states $|\psi_{a\tau}\rangle_\sm$ are Gaussian
wave-packets centered $a$, with zero 'momentum' and a fixed width $\tau$.
We also assume that the {\em a priori} probability $p(a)$ of
the symbol $a$, {\em i.e.} the probability to have a signal
centered in $a$, is given by a Gaussian $$p(a) =
\frac{1}{\sqrt{2\pi}\Delta} \exp\left(-\frac{a^2}{2\Delta^2}
\right)$$ of zero mean and width $\Delta$.
The width $\Delta$ will be referred to as the {\em size} of the
transmitted alphabet.
Notice that the class $\{|\psi_{a\tau}\rangle_\sm\}$  is made by
non-orthogonal states, we have.
$$
|{}_{\sm}\langle\psi_{b\tau'}|\psi_{a\tau}\rangle_\sm|^2
= \frac{2\tau\tau'}{\tau^2+\tau'^2}\exp\left\{-
\frac{(a -b)^2}{2(\tau^2+\tau'^2)} \right\}
\:,$$
and, in particular,
$$
|{}_{\sm}\langle\psi_{b\tau}|\psi_{a\tau}\rangle_\sm|^2
= \exp\left\{-\frac{(a -b)^2}{4\tau^2} \right\}
\:.$$
Upon defining the standard dual basis, say momentum eigenstates,
as
\begin{eqnarray}
|p\rangle = \frac{1}{\sqrt{2\pi}}\int dx\: e^{ipx}|x\rangle
\end{eqnarray}
one has that the position- and momentum-like observables are given by
$$X=\int dx \:x \: |x\rangle\langle x| \quad P=\int dp \:p
\: |p\rangle\langle p|\:,$$
whereas the energy operator reads as follows
$N=\frac12 (X^2 + P^2)$.
The average energy
$N_\sm(a) ={}_{\sm}\langle\psi_{a\tau}|N|\psi_{a\tau}\rangle_{\sm}$
of the signal
is thus given by
\begin{eqnarray}
N_\sm(a) = \frac{1}{2}\left(a^2 + \tau^2 + \frac{1}{4\tau^2}  \right) \:.
\nonumber
\end{eqnarray}
Finally, the mean energy sent into the channel {\em per use}
(from now on the {\em signal energy}) reads as follows
\begin{eqnarray}
 N_\sm =
 \int da \: p(a)N_\sm(a) = \frac{1}{2}\left( \Delta^2 + \tau^2 + \frac{1}{4\tau^2}
\right). \nonumber
\end{eqnarray}
For each signal $|\psi_{a\tau}\rangle_{\sm}$ we have $\Delta X^2 =\tau^2$
and $\Delta P^2= (4\tau^2)^{-1}$ that is, the signals
$|\psi_{a\tau}\rangle_{\sm}$ are minimum uncertainty states. For
$\tau^2=1/2$ one has equal variances, whereas $\tau^2 \neq 1/2$
corresponds to "squeezing" of the signals.
The signal energy is minimum for $\tau^2 =1/2$; in this case we
have $N_\sm = \frac12 (1+\Delta^2)$.
\par
Notice that transmitted symbols may be viewed as shift parameters
$|\psi_{a\tau}\rangle_\sm = \exp(-i P a)|\psi_{0\tau}\rangle_{\sm}$
imposed to the 'undisplaced' basic state $|\psi_{0\tau}\rangle_{\sm}$.
This feature will be used in optimizing the measurement at the output.
\subsection{Preparation of the probe state}
\begin{figure}[h]
\includegraphics[width=0.4\textwidth]{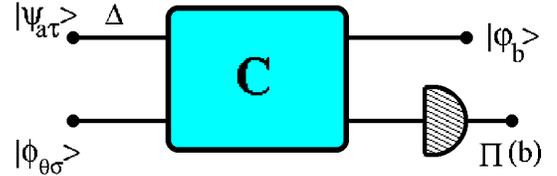}
\caption{Schematic diagram of an indirect measurement scheme for continuous
variable systems. The symbols from the input alphabet ${\mathbbm
A}$ are encoded in a set of Gaussian signals
$|\psi_{a\tau}\rangle_\sm$ with real amplitude $a$ and fixed width
$\tau$. The {\em a priori} probability of the transmitted symbols
is a Gaussian with zero mean and width $\Delta$. The signal is
coupled, by a $C_{sum}$ gate, to a probe excited in the known
state $|\phi_{\theta\sigma}\rangle_\pm$. After the interaction the
probe is measured in order to infer the transmitted state, while
the signal is left for the subsequent user. The measurement is
described by the operator-valued measure $\Pi(b)$, which is
optimized in order to maximize the estimation fidelity. The output
signals $|\varphi_b\rangle_\sm$ are the conditional states of the
signal after having observed the outcome $b$.
\label{f:QrepScheme}}
\end{figure}

The setup of the measurement scheme is shown in Fig.
\ref{f:QrepScheme}. The signal is coupled with a probe system
excited in the (known) state
\begin{align}
|\phi_{\theta\sigma}\rangle_\pm &= \cos\theta
\int dx\: g_{0,\sigma}(x)\:|x\rangle_\pm +
\gamma\sin\theta\int dp\: g_{0,\sigma}(p)\: |p\rangle_\pm
\nonumber \\
&= \int dx\: \left[\cos\theta\: g_{0,\sigma} (x) +
\gamma\sin\theta \: g_{0,(2\sigma)^{-1}}(x) \right] \: |x\rangle_\pm
\label{probe}
\end{align}
where $\theta \in [0,\pi/2]$,
and
\begin{eqnarray}
\gamma=\frac{\sqrt{1+\beta^2\tan^2\theta}-1}{\beta\tan\theta}\qquad \beta^2 =
\sigma^2 + \frac{1}{4\sigma^2}
\end{eqnarray}
is a normalization factor.
\par
The state
$|\phi_{\theta\sigma}\rangle_\pm$,
in close analogy with the finite-dimensional case
\cite{Fid}, is built as a tunable superposition of the
\emph{almost localized} state (up to the width $\sigma$)
$|\phi_{0\sigma}\rangle_\pm$
and the \emph{almost delocalized} state
$|\phi_{0(2\sigma)^{-1}}\rangle_\pm$.
The probe state depends on two parameters: $\theta$
and the width $\sigma$. This apparent redundancy can be eliminated
upon imposing a constraint on the probe energy $N_\pm(\sigma,\theta)
={}_{\pm}\langle\phi_{\theta\sigma}|N|\phi_{\theta\sigma}\rangle_\pm$,
whose expression reads as follows
\begin{align}
N_\pm(\beta,\theta) &= \frac{1}{2}\left[\beta^2 \left(\cos^2\theta
+ \gamma^2\sin^2\theta\right)
+\frac{\gamma}{\beta^3} \sin2\theta
\right] \nonumber \\
&= \frac12 \left[ \beta^2 +
\frac{2\cos^2\theta(\beta^4-1)}{\beta^4}\left(1-\sqrt{1+\beta^2\tan^2\theta}
\right)\right]
\end{align}
At a fixed value of $\theta$ the probe energy is minimum for
$\sigma^2=1/2$ ($\beta=1$), corresponding to $N_\pm\left(1, \theta
\right) =\frac{1}{2}$, whereas at a fixed value of $\sigma$ the
probe energy is minimum for
\begin{eqnarray}
\tan^2 \frac{\theta}{2} = 1+ 2\: \frac{\beta-
\sqrt{2\beta(1+\beta)}}{2+\beta}
\end{eqnarray}
corresponding to $N_{\pm} (\beta)=
\frac{1+\beta(\beta-1)(\beta^2+1)}{2\beta^2}$.\\
The two-parameter nature of the probe signal will be used to
analyze the information-disturbance trade-off in different
configurations, which include regimes at fixed energy as well as
regime with increasing energy.
\subsection{Interaction}
The signal and the probe are then coupled by a $C_{sum}=\exp\{ -i
X_{\sm} \otimes P_{\pm}\}$ gate
(denoted by ${\mathbf C}$ in Fig. \ref{f:QrepScheme}), which acts
on the standard basis of the signal-probe system as follows
$$ C_{sum}|x\rangle_\sm|y\rangle_\pm = |x\rangle_\sm|x+y\rangle_\pm $$
{\em i.e.} represents the generalization to continuous variable
of the $C_{not}$ gate. \cite{CVOperations}
The global (entangled) signal-probe state
$|\psi_{out}\rangle\rangle_{\sm\pm}$
after the interaction is given by
\begin{widetext}
\begin{align}
|\psi_{out}\rangle\rangle_{\sm\pm} &= {\mathbf C}|\psi_{a\tau}\rangle_\sm
\otimes |\phi_{\theta\sigma}\rangle_\pm \nonumber \\
&= \cos\theta\int\!\!\!\int dx\:dy\:
g_{\sigma,0}(x)g_{a,\tau}(y)|y\rangle_\sm\otimes |x+y\rangle_\pm
+ \: \frac{\gamma\sin\theta}{\sqrt{2\pi}} \int\!\!\!\int\!\!\!\int\! dp\:dx'\: dy\:
g_{0,\sigma}(p)\:g_{a,\tau}(y)\:e^{ipx'}|y\rangle_\sm\otimes|x'+y\rangle_\pm
\nonumber \\
&= \int\!\!\!\int dx\:dy\: g_{a,\tau}(y) \left[\cos\theta
g_{0,\sigma}(x)+\gamma\sin\theta g_{0,(2\sigma)^{-1}}(x) \right]
|y\rangle_\sm\otimes|x+y\rangle_\pm \:.
\end{align}
\end{widetext}
\subsection{Measurement}
After the interaction the probe is measured in order to infer the
transmitted state, while the signal is left for the subsequent user.
The measurement is described by the operator-valued measure $P(b)=I\otimes
\Pi(b)$, where $\Pi(b)$ is an operator-valued measure acting on the sole probe
Hilbert space. Since the transmitted symbols are Gaussian
distributed shift parameters we expect the optimal measurement to be of the
form \cite{hel}
$$\Pi(b)=\frac{1}{\kappa}\: |b/\kappa\rangle_\pm{}_{\pm}\langle b/\kappa|$$
with $|b\rangle_\pm$ being standard
basis states (position eigenstates) and $\kappa$ a real constant,
hereafter referred to as the measurement gain, chosen
to optimize the desired figure of merit (here the estimation fidelity).
In order to estimate the transmitted state from the outcomes of the
measurement we assume the natural inference rule
$$ b \rightarrow |\psi_{b\tau}\rangle_\sm$$  where $|\psi_{b\tau}\rangle_\sm$
is of the form (\ref{signal}) {\em i.e.} a signal Gaussian wave-packet
centered in $b$ and width $\tau$.
The probability density $q(b)$ of obtaining the outcome "$b$", and
the expression of the corresponding conditional state $\varrho_b$
for the signal are thus given by
 \begin{widetext}
\begin{align}
q(b) &= \hbox{Tr}_{\sm\pm}\left[|\psi_{out}\rangle\rangle_{\sm\pm}
{}_{\sm\pm}\langle\langle \psi_{out}| \:
{\mathbbm I} \otimes \Pi(b)\right]   \label{c1}
= {}_\sm\langle \tilde\varphi_b| \tilde\varphi_b \rangle_\sm
\\
\varrho_b&=
\frac{1}{q(b)}\hbox{Tr}_{\pm}\left[|\psi_{out}\rangle\rangle_{\sm\pm}
{}_{\sm\pm}\langle\langle \psi_{out}| \:
{\mathbbm I} \otimes \Pi(b)\right] \label{c2}
= \frac{1}{q(b)} | \tilde\varphi_b \rangle_\sm{}_\sm\langle \tilde\varphi_b| =
|\varphi_b\rangle_\sm {}_\sm\langle \varphi_b |
\end{align}
The last equalities in both Eqs. (\ref{c1}) and (\ref{c2}), which
express the purity of the conditional state,
follow from the fact that the initial states of both the signal
and the probe are pure, and that the measure $\Pi(b)$ is pure too.
Mixed measurements may be considered as well, though unavoidably leading
to additional {\em extrinsic} noise \cite{hel,hol}.
The unnormalized signal states $|\tilde{\varphi}_b\rangle_\sm$
are given by
 \begin{align}
 |\tilde{\varphi}_b\rangle_\sm = \frac{1}{\sqrt{\kappa}}{}_{\pm}\langle
 b|\psi_{out}\rangle\rangle_{\sm\pm} =& \frac{\cos\theta}{\sqrt{\kappa}}\int\!\!\!\int dx\:dy\:
         g_{\sigma,0}(x)\:g_{a,\tau}(y)\:\delta(b/\kappa-x-y)\:|y\rangle_\sm \nonumber \\
 & + \frac{\gamma\sin\theta}{\sqrt{2\pi\kappa}}\int\!\!\!\int\!\!\!\int dp\:dx'\: dy\:
 g_{0,\sigma}(p)\:g_{\sigma,a}(y)\:e^{ipx'}\:\delta(b/\kappa-x'-y)
 \:|y\rangle_\sm
 \nonumber \\ =& \int \frac{dy}{\sqrt{\kappa}}\: g_{a,\tau}(y)
 \left[\cos\theta g_{b/\kappa,\sigma}(y) + \gamma\sin\theta\:
 g_{b/\kappa,(2\sigma)^{-1}}(y) \right]|y\rangle_\sm
 \label{c3}
 \end{align}
 thus leading to
\begin{eqnarray}
q(b) = \frac{1}{\kappa}\int dy\: |g_{a,\tau}(y)|^2\: \left[\cos\theta
g_{b/\kappa,\sigma}(y) + \gamma\sin\theta\: g_{b/\kappa,(2\sigma)^{-1}}(y)
\right]^2
\label{pz}\:.
\end{eqnarray}
In Eqs. (\ref{c1}), (\ref{c2}) and (\ref{c3}), for the sake of
a simpler notation, we omitted the explicit dependence of the conditional
states on the signal and probe widths, $\tau$ and $\sigma$.
\section{Information-disturbance trade-off}\label{s:fid}
We are now in the position of evaluating the fidelities.
As concern the transmission fidelity, according to Eq.
(\ref{F_psi}) we have
$$ F_a = \int db\: q(b) |{}_{\sm}\langle\varphi_b|\psi_{a\tau}\rangle_\sm|^2
= \int db\: |{}_{\sm}\langle\tilde{\varphi}_b|\psi_{a\tau}\rangle_{\sm}|^2
$$
After lengthy but straightforward calculations one arrives at
\begin{align}
F_a &= \frac{\sqrt{2}\sigma\cos^2\theta}{\sqrt{2\sigma^2 + \tau^2}} +
\frac{4\sigma\cos\theta\left(-2\sigma\cos\theta + \sqrt{1 + 4\sigma^4 -
(1-2\sigma^2)^2\cos^2\theta} \right)}{\sqrt{(1+4\sigma^4)(1 + 4\sigma^4 +
4\sigma^2\tau^2)}} + \nonumber \\
&+ \frac{\left(-2\sigma\cos\theta + \sqrt{1+4\sigma^4 - (1-2\sigma^2)^2\cos^2
\theta}\right)^2}{(1+4\sigma^4)\sqrt{1+2\sigma^2\tau^2}}\:.
\label{avF}
\end{align}
Notice that $F_a$ does not depends on the amplitude
$a$, nor on the measurement gain $\kappa$.
Therefore the average fidelity $F$
$$ F = \int da\: p(a)\:F_a = F_a \:,$$
is equal to the signal fidelity and also does not depends on the
alphabet size $\Delta$.
\par
Using Eqs. (\ref{G_psi}) and (\ref{pz}) one evaluates the estimation fidelity
$G_a$ as follows
\begin{eqnarray} G_a =
\int db\: q(b)\: |{}_{\sm}\langle\psi_{b\tau}|\psi_{a\tau}\rangle_{\sm}|^2 \:.
\end{eqnarray}
The signal fidelity $G_a$ depends on the amplitude $a$ of the
signal, by averaging  over the \emph{a priori} signal probability
$p(a)$ we arrive at the estimation fidelity $G$
\begin{align}
\nonumber
G &= \int da\: p(a)\:G_a \\ &= \frac{\sqrt{2}\tau\cos^2\theta
}{\sqrt{\Delta^2(\kappa -1)^2 + 2\tau^2 + \kappa^2(\sigma^2 + \tau^2)}}
- \frac{\sqrt{2}\tau\cos^2\theta \left(8\sigma^2 -
4\sigma\sqrt{4\sigma^2 + (1 +
4\sigma^4)\tan^2\theta}\right)}{\sqrt{(1+4\sigma^4)(\Delta^2(\kappa-1)^2(1+4\sigma^4)
+ 2(1+4\sigma^4)\tau^2 + \kappa^2(2\sigma^2 + \tau^2 + 4\sigma^4\tau^2))}} +
\nonumber \\ & + \frac{\sqrt{2}\tau\cos^2\theta \left[16\sigma^3 + 2(\sigma +
4\sigma^5)\tan^2\theta - 8\sigma^2\sqrt{4\sigma^2 +
(1+4\sigma^4)\tan^2\theta}\right]}{(1+4\sigma^4)\sqrt{4\sigma^2(\Delta^2
+2\tau^2) + \kappa^2(1 + 4\Delta^2\sigma^2 + 4\sigma^2\tau^2) -
8\kappa\Delta^2\sigma^2}}  \:,
\label{avG}
\end{align}
\end{widetext}
which, besides the signal and probe  widths, depends on the
alphabet size $\Delta$ and the measurement gain $\kappa$.
\par
\begin{figure}[h]
\begin{tabular}{cc}
\includegraphics[width=0.23\textwidth]{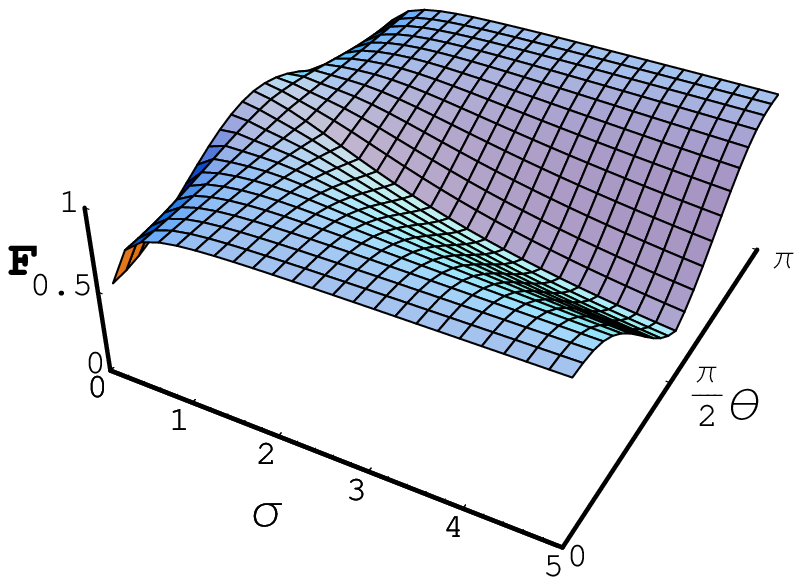} &
\includegraphics[width=0.23\textwidth]{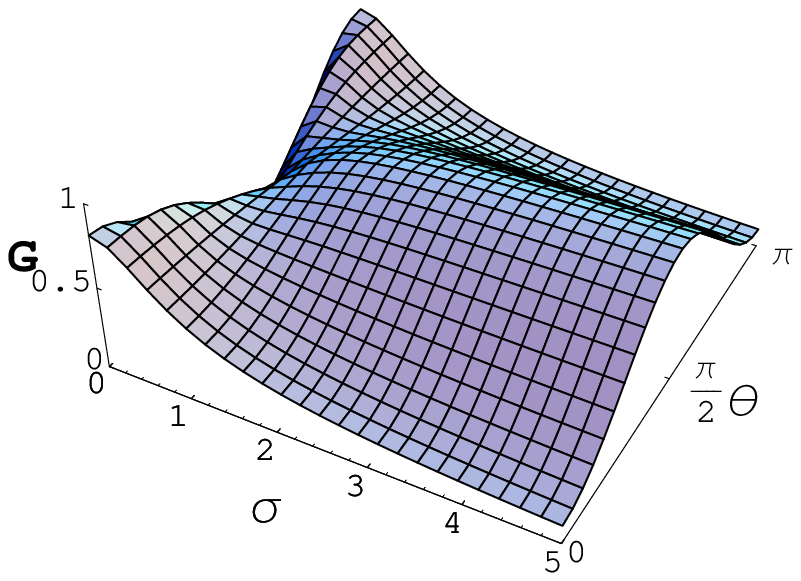} \\
\includegraphics[width=0.23\textwidth]{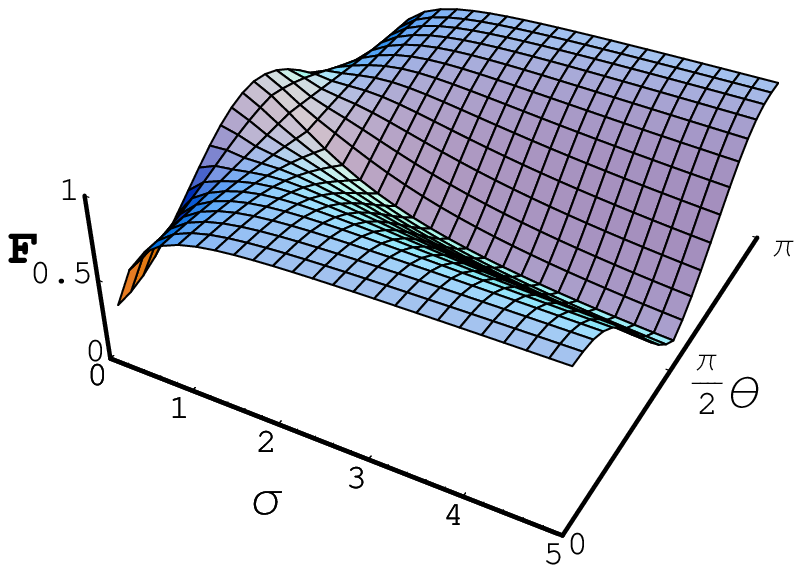} &
\includegraphics[width=0.23\textwidth]{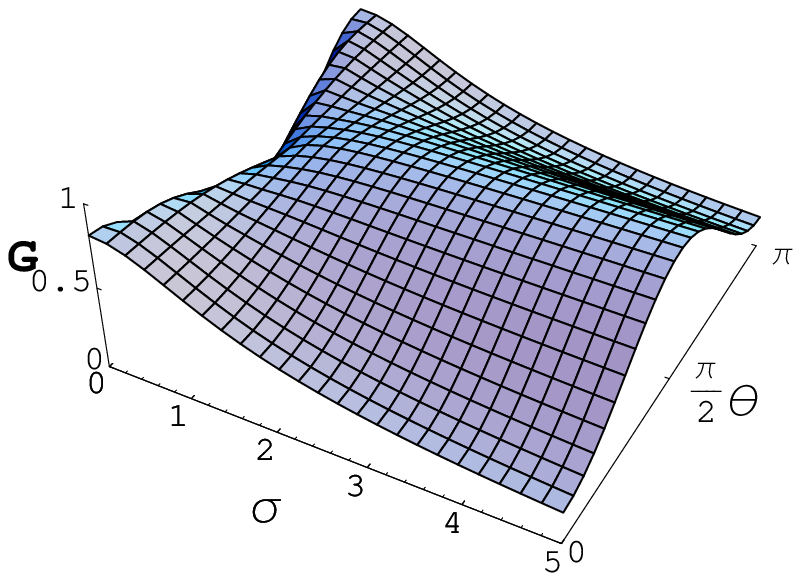} \\
\includegraphics[width=0.23\textwidth]{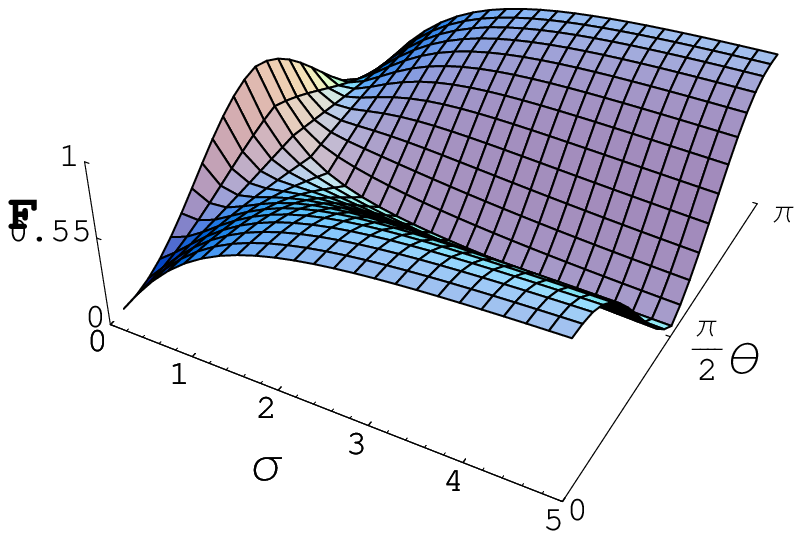} &
\includegraphics[width=0.23\textwidth]{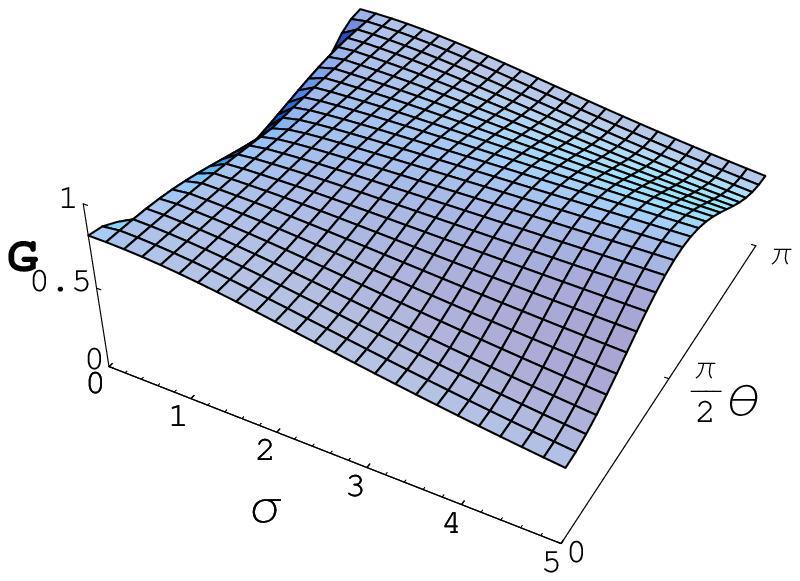}
\end{tabular}
\caption{Left: Transmission fidelity $F(\sigma,\theta)$ for $\kappa=1$ and
for different values of $\tau$; from top to bottom $\tau=
0.4,1/\sqrt{2},2$.
Right: Estimation fidelity $G(\sigma,\theta)$ for $\kappa=1$,
$\Delta=1/\sqrt{2}$  and for different values of $\tau$;
from top to bottom $\tau=0.4,1\sqrt{2}, 2$.  } \label{f:Fids3D}
\end{figure}

By inspecting Eqs. (\ref{avF}) and (\ref{avG})
the superposition nature of the probe state $|\phi_{\theta\sigma}\rangle_\pm$
can be clearly seen: at fixed values of the parameters $\kappa$ and
$\Delta$ the fidelities oscillate as a function of the tuning parameter
$\theta$ (see Fig. \ref{f:Fids3D}). As it is apparent from the plots
for $\sigma^2<1/2$ the transmission (estimation) fidelity $F$ ($G$)
is maximized (minimized) at $\theta=\pi/2$ and
is minimized (maximized) at $\theta=0$, whereas for $\sigma^2>1/2$
the situation is the opposite.
By varying the values of $\kappa$ and $\Delta$ the shape
of the curves slightly change, while the overall behaviour
is the same.
\par
In order to derive a proper information/disturbance
trade-off we have first to optimize (maximize) the estimation
fidelity with respect to the measurement gain $\kappa$. The general solution
of optimization equation is rather involved and does not offer a
clear picture. Therefore, in order to gain insight on the general
behaviour, we now proceed to analyze the optimization
and the corresponding trade-offs for relevant configurations.
\subsection{Probe in the high-energy limit}
\begin{figure}[h]
\includegraphics[width=0.4\textwidth]{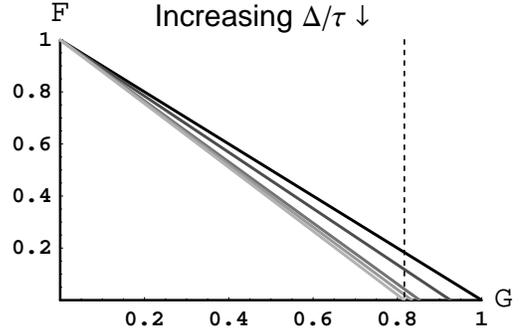}
\caption{Information/disturbance trade-off $F_A(G,y)$ in the
high energy limit for the probe, for different values of the
width ratio
$y=\Delta^2/\tau^2$. From darker to lighter gray
we plot the trade-off for
$y=0.5,3,7,10000$. The trade-off get worse for increasing
values of $y$. For $y\gg 1$ the function $F_A(G,y)$ intercepts the
axis $F=0$ for $G=\sqrt{2/3}$. One moves along the curve, from right
to left, by increasing $\theta$.} \label{f:PlotA}
\end{figure}

In order to compare the optimal scheme for qudit
to the present CV scheme we
start by considering the probe in the state
\begin{equation}
|\psi_{\theta 0}\rangle = \cos\theta|0\rangle_\pm +
\frac{\sin\theta}{\sqrt{2\pi}}\int dx\: |x\rangle_\pm .
\end{equation}
which is the plain analogue of the probe used in
the optimal scheme for qudit \cite{Fid}.
We obtain this configuration by taking the limit $\sigma\rightarrow
0$ in (\ref{probe}). Of course this is an ideal case, since it corresponds to
a probe state with divergent energy.
As concerns the optimization of the measurement we
obtain
\begin{equation}
\kappa_{opt} = \frac{\Delta^2}{\Delta^2 + \tau^2}
\end{equation}
which corresponds to fidelities
\begin{eqnarray}
F &=& \sin^2\theta  \\
G &=& \cos^2\theta\sqrt{\frac{1 + \frac{\Delta^2}{\tau^2}}{1 +
\frac{3}{2}\frac{\Delta^2}{\tau^2}}}
\end{eqnarray}
and to the parametric function $F=F_A(G,y)$:
\begin{equation}
F_{\textbf{A}}(G,y) = 1 - G\sqrt{\frac{1 + \frac{3}{2}y}{1 + y}}
\: \: \: \: \: \: \: \:  y=\frac{\Delta^2}{\tau^2} \label{FA}
\end{equation}
which depends on the ratio $y$ between the alphabet size and the
signal width.
We have a linear dependence between the two fidelities and for
each curve one can explore the trade-off by varying the parameter
$\theta$: one moves along the curve from right to
left by increasing $\theta$. Different curves for different values
of the ratio $y$ are depicted in Fig. \ref{f:PlotA}.
We see that the high-fidelity region (both $F$ and $G$ close to
unity) is excluded and that the
trade-off is better for small values of the ratio $y$,
{\em i.e.} for "small alphabets".
For increasing $y$, {\em i.e.} for increasing size of the
alphabet, the slope of the curve $F_{\textbf{A}}(G,y)$ decreases.
The function $F_{\textbf{A}}(G,y)$
intercepts the \emph{G-axis} at $G=\sqrt{(1+y)/(1+\frac32 y)}$.
\subsection{Probe in the undisplaced state}
Here we analyze the case of undisplaced probe state
$|\phi_{0\sigma}\rangle_\pm$ which, in the limit
$\sigma\rightarrow 0$ approaches the localized state
$|0\rangle_\pm$. We can
obtain this configuration by setting $\theta=0$ in Eq. (\ref{probe}).
Maximizing the estimation fidelity with respect to
the measurement gain we arrive to
$$
\kappa_{opt} = \frac{\Delta^2}{\Delta^2 + \tau^2 + \sigma^2}
$$
and, correspondingly, to the fidelities
\begin{eqnarray}
F &=& \sqrt{\frac{2z}{1 +
2z}} \hspace{2.8cm} z=\frac{\sigma^2}{\tau^2} \\
G &=& \sqrt{\frac{1 + z + y}{1 + z + \frac{y}{2}\left(3 + z
\right)}} \label{Gth0}
\:,
\end{eqnarray}
which, besides the ratio $y$, also depend on the ratio $z$ between
the probe and the signal widths.
\par
\begin{figure}[h]
\includegraphics[width=0.4\textwidth]{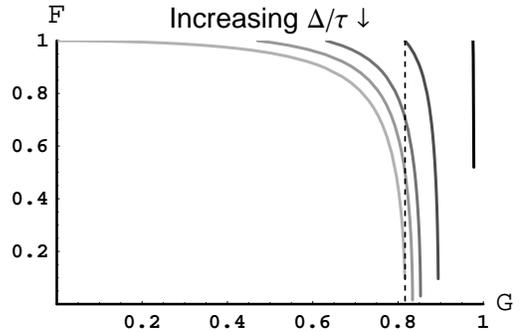}
\caption{Information/disturbance trade-off $F_B(G,y)$ for the probe in
the undisplaced state $|\psi_{0\sigma}\rangle_\pm$ and for
different values of the ratio $y=\Delta^2/\tau^2$.
From darker to lighter gray:
the trade-off for $y=0.1,1,3,7,10000$. The trade-off get worse for
increasing values of $y$. For $y\gg 1$ the function $F_B(G,y)$
intercepts the {\em G-axis} at $G=\sqrt{2/3}$. One moves along the
curves from right to left by increasing $\sigma$.} \label{f:PlotB}
\end{figure}

Upon inverting Eq. (\ref{Gth0}) we arrive at the parametric function
$F=F_B(G,y)$, which depends only on the ratio $y$:
\begin{equation}
F_{\textbf{B}}(G,y) = \sqrt{\frac{G^2\left(4 + 6y\right) -4\left(1
+ y\right)}{G^2\left(2 + 5y\right) -4\left(\frac{1}{2} +
y\right)}} \label{FB}
\end{equation}
In Fig. \ref{f:PlotB} we show the trade-off for
different values of $y$.
\par
As it is apparent from the plot, this configuration allows
to achieve the high-fidelity region. The trade-off is
worse for larger values of the ratio between the alphabet
size and the signal width. Therefore, in order to get
superior performances, it is preferable to have a
small alphabet rather than a class of narrow signals.
For fixed width of the signals this is intuitively
expected: the larger is the alphabet the worse is the trade-off.
On the other hand, for a fixed alphabet size, this means
that the larger are the signals the better is the trade-off.
One moves along each curve by tuning the probe width
$\sigma$: from right to left by increasing $\sigma$.
\subsection{Probe in the minimum energy state}
\begin{figure}[h]
\includegraphics[width=0.4\textwidth]{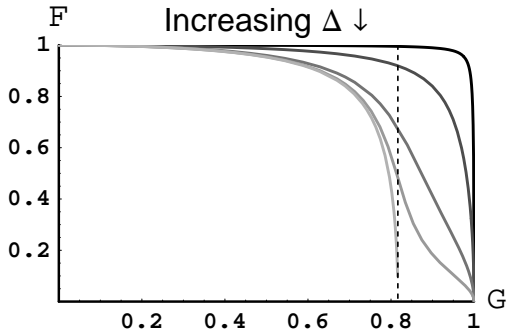}
\caption{Information/disturbance trade-off $F_C(G,\Delta)$ for
the probe with minimum energy and for different values of
the alphabet width $\Delta$. From darker to lighter gray:
the trade-off for $\Delta=0.1,0.2,2,5,\infty$. The trade-off get
worse for increasing values the $\Delta$. For $\Delta \rightarrow \infty$
the  function $F_C(G,\Delta)$
intercepts the {\em G-axis} at $G=\sqrt{2/3}$. One moves along
the curves from left to right by increasing $\tau$.} \label{f:PlotC}
\end{figure}

As we have already seen, at fixed $\theta$ we have minimum
energy for $\sigma^2=1/2$. In this case we also lose the dependency on
$\theta$ and the probe state is given by
\begin{equation}
|\phi_{\theta2^{-1/2}}\rangle_\pm = \int dx\: g_{0,2^{-1/2}}(x)
\:|x\rangle_\pm \nonumber
\end{equation}
The optimal $\kappa$ is given by
\begin{equation}
\kappa_{opt} = \frac{2\Delta^2}{1 + 2\Delta^2 + 2\tau^2},
\end{equation}
which corresponds to fidelities
\begin{eqnarray}
F &=& \sqrt{\frac{1}{1 + \tau^2}}  \\
G &=& \tau\sqrt{\frac{2(1 + 2\Delta^2 + 2\tau^2)}{4\tau^4 +
\Delta^2 + \tau^2(1+3\Delta^2)}}
\end{eqnarray}
and to the parametric function
\begin{widetext}
\begin{equation}
F_{\textbf{C}}(G,\Delta)= \sqrt{\frac{3 - 2\Delta^2
+3G^2(\Delta^2-1) - \sqrt{(1+2\Delta^2)^2
-2G^2(1+3\Delta^2+6\Delta^4)+G^4(1+2\Delta^2 +
9\Delta^4)}}{2-4\Delta^2 - G^2(5\Delta^2 -2)}}. \label{FC}
\end{equation}
\end{widetext}
The trade-off $F_{\textbf{C}}(G,\Delta)$, which depends only
on the alphabet size $\Delta$, is shown for different values
of $\Delta$ in Fig. \ref{f:PlotC}.
\par
Also this configuration permits to access the high-fidelity
region. The curves corresponding to smaller values of $\Delta$ are the
upper ones {\em i.e.} the trade-off is worse for larger
alphabets. One moves along the curves by tuning the signal width
$\tau$: from left to right by increasing $\tau$. For narrower
signal we have less disturbances, though we get less information too.
In the limit for $\Delta\rightarrow 0$ we have $F_C(G,0)=1$, in
particular, up to the second order in $\Delta$ (see
Fig.\ref{f:PlotC}, upper curve) we have
\begin{equation}
F_C(G,\Delta) = 1 + \frac{G^2}{4(G^2-1)}\Delta^2 + O(\Delta^4).
\end{equation}
For uniform alphabets, {\em i.e.} in the limit
$\Delta\rightarrow\infty$ Eq. (\ref{FC}) rewrites as
\begin{equation}
F = \sqrt{\frac{4 - 6G^2}{4-5G^2}} \label{Bound}
\end{equation}
We may assume Eq.(\ref{Bound}) as the CV bound for signals chosen
from a flat distribution in the Hilbert space. This should be
compared with the rigorous bound derived in \cite{KB} for random
qudits,  \emph{i.e.} for signals uniformly distributed
in a d-dimensional Hilbert space
\begin{equation}
F = \frac{1}{d+1} + \left( \sqrt{G - \frac{1}{d+1}} +
\sqrt{(d-1)\left(\frac{2}{d+1} -G\right)}\right)^2
\:.\end{equation}
\par
A continuous-variable system thus appears to offer the possibility
of a superior trade-off at the price of increasing the energy
impinged into the device. Notice, however, that increasing the
probe energy does not necessarily lead to better trade-off:
{\em e.g.} compare Eq.(\ref{FA}) to Eqs.
(\ref{FB}) and (\ref{FC}). On the other hand, the trade-off
strongly depends on the ratio between the alphabet size and the signal
width, which in turn determine the allocation of the signal
energy.
\subsection{Comparison between different probe configurations}
\begin{figure}[h]
\includegraphics[width=0.4\textwidth]{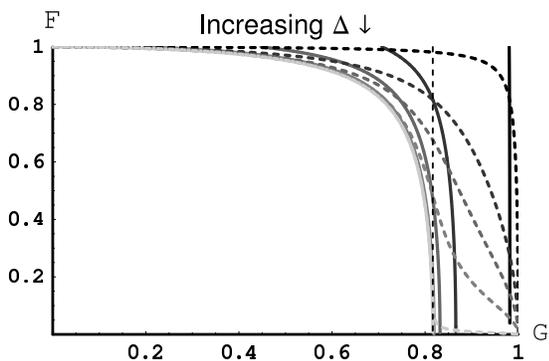}
\caption{Information/disturbance trade-offs $F_B(G,2\Delta^2)$ and
$F_C(G,\Delta)$ (dashed lines) for different values of $\Delta$.
From darker to lighter gray: the trade-offs for
$\Delta=0.2,1,2,5,100$. For $\Delta\gtrsim 3$ the two
configurations lead to similar results. For $\Delta<3$ configuration
\textbf{B} favors the information fidelity while configuration
\textbf{C} advantages the estimation fidelity.} \label{f:PlotBC}
\end{figure}

Here we compare the trade-offs achievable by setting
the probe in the undisplaced state $|\phi_{0\sigma}\rangle_\pm $
(configuration \textbf{B}) and in the minimum energy state
$|\phi_{\theta2^{-1/2}}\rangle_\pm$ (configuration \textbf{C})
respectively.
Since (\ref{FB}) depends on the ratio $\Delta/\tau$ and
(\ref{FC}) depends only on $\Delta$ we compare the two configurations
by fixing the signal width. The value $\tau^2=1/2$ has been chosen
in order to have signals with minimum energy. The trade-off
$F_{\textbf{B}}(G,y)$ rewrites as
\begin{equation}
F=F_{\textbf{B}}(G,2\Delta^2)= \sqrt{\frac{2(G^2(3\Delta^2 -1) + 1
-2\Delta^2)}{G^2(5\Delta^2 +1) - 4\Delta^2 -1}} \nonumber
\end{equation}
In Fig. \ref{f:PlotBC} the trade-offs are compared
for different values of $\Delta$.
\par
For alphabet sizes larger than a threshold $\Delta\gtrsim \Delta_{th}$
the curves are very similar, approximately leading to the same
trade-off. On the other hand, for $\Delta<\Delta_{th}$ configuration
\textbf{B} favors the information fidelity while configuration
\textbf{C} advantages the estimation fidelity. For $\tau^2=1/2$ we
have roughly $\Delta_{th}\simeq 3$.
\section{Optical implementation}\label{s:imp}
In this section we briefly mention how the building blocks of our
scheme can be implemented in a quantum optical scenario. The main
goal is show that the present scheme may be implemented with currently
technology, rather than describing a specific measurement scheme.
\par
The measurement in the standard basis corresponds to homodyne 
detection, whereas the signals $|\psi_{a\tau}\rangle$ are Gaussian 
wavepackets with amplitude
$a$ and width $\tau$. They correspond to squeezed-coherent states
of the form $D(a/\sqrt{2})S(r)|0\rangle$ where $|0\rangle$ is the 
em vacuum, $D(\alpha)=\exp(\alpha a^\dag - \bar\alpha a)$ is the 
displacement operator and $S(r)=\exp[\frac12(\zeta^2 a^{\dag 2} - \bar\zeta^2
a^2) ]$ is the squeezing operator.
In order to obtain the signals $|\psi_{a\tau}\rangle$, 
the relation $\sinh^2 r = \frac12 (\tau^2 + \frac{1}{4\tau^2})$ 
must hold. 
As concerns the interaction: the C-sum gate $C_{sum}=
\exp\{ -i X_{\sm} \otimes P_{\pm}\}$
expressed in terms of the mode operators reads as follows 
$C_{sum}=\exp\{\frac12 (a^\dag b^\dag + a b^\dag - ab - a^\dag b)\}$
and may be implemented by parametric interactions in 
second order $\chi^{(2)}$ nonlinear crystals. Alternatively,
conditional schemes involving both linear and nonlinear 
interactions have been also proposed \cite{msa,xqn}.
As a matter of $C_{sum}$ interaction between light pulses has been 
investigated in several previous 
experiments \cite{ls1}. In addition, it has
been experimentally observed between the polarization of
light pulses and collective spin of huge atomic 
samples \cite{ls2}. 
\section{Conclusions}\label{s:out}
We have suggested a class of indirect measurement schemes to
estimate the state of a
continuous variable Gaussian system without destroying the 
state itself. The schemes involve a single
additional probe and allow for the nondemolitive transmission of
a continuous real alphabet over a quantum channel.  The trade-off
between information gain and state disturbance has been
quantified by fidelities and optimized with respect to the
measurement performed after the signal-probe interaction.  Different
configurations have been analyzed in terms of the energy carried
by the signal and the probe.  A bound for a class of randomly
distributed CV signals has been derived, which may be compared
with the analogous (general) bound derived for qudits \cite{KB}.
We found that a continuous-variable system generally offers the
possibility of a better trade-off at the price of increasing the
overall energy introduced into the device. Notice, however, that
increasing the probe energy does not necessarily lead to a better
trade-off, the most relevant parameter being the ratio between
the alphabet size and the signal width, which in turn determine the
allocation of the signal energy.
\section*{Acknowledgments}
MGAP thanks N. Cerf for discussions. This work has been
supported by MIUR through the project PRIN-2005024254-002.


\begin{thebibliography}{99}
\bibitem{hh0} H. F. Hofmann, Phys. Rev. A {\bf 62}, 022103 (2000).
\bibitem{nocl} W. K. Wootters and W. K.Zurek, Nature (London)
{\bf 299},
802 (1982); V. Buzek, M. Hillery, Phys. Rev. A {\bf 54}, 1844
(1996); N. Gisin, S. Massar Phys. Rev. Lett. {\bf 79}, 2153
(1997); R. F. Werner, Phys. Rev. A {\bf 58}, 1827 (1998).
\bibitem{telecl} M. Murao et al., Phys. Rev. A {\bf 59}, 156
(1999); P.
van Loock, S. Braunstein, Phys Rev Lett. {\bf 87}, 247901
(2001);
A. Ferraro et al., J. Opt. Soc. Am. B {\bf 21}, 1241 (2004).
\bibitem{KB} K. Banaszek, Phys. Rev. Lett. {\bf 86}, 1366
(2001).
\bibitem{gra} J. F. Roch, J. P. Poizat, P. Grangier,
Phys. Rev. Lett. {\bf 71}, 2006 (1993).
\bibitem{hau} Y. Yamamoto, Science {\bf 263}, 1394 (1994)
\bibitem{briegel} H. J. Briegel,W. Dür,J. J. Cirac and P. Zoller, 
Phys. Rev. Lett. {\bf 81}, 5932 (1998).
\bibitem{child1} L. Childress, J.M. Taylor, A. S. Sørensen and M.D Lukin, preprint ArXiv: quant-ph/0410123.
\bibitem{child2} L. Childress, J.M. Taylor, A. S. Sørensen and M.D Lukin, preprint ArXiv: quant-ph/0502122.
\bibitem{kok} P. Kok, C. P. Williams and J. P. Dowling, Phys. 
Rev. A {\bf 68}, 022301 (2003).
\bibitem{shi} B.-S. Shi,Y.-K. Jiang, G.-C. Guo, Physica A {\bf 284}, 107 (2000).
\bibitem{Fid} M. G. Genoni, M. G. A. Paris, 
Phys. Rev. A {\bf 71}, 052307  (2005).
\bibitem{hh1} H. F. Hofmann, Phys. Rev. A {\bf 67}, 022106 (2003).
\bibitem{rmp} C. M. Caves, P. D. Drummond, Rev. Mod. Phys. {\bf 66},
481537 (1994).
\bibitem{sze} M. G. A. Paris, Fortschr. Phys. {\bf 51}, 202 (2003).
\bibitem{hel} C.W.Helstrom, {\em Quantum Detection and
Estimation Theory} (Academic Press, New York, 1976).
\bibitem{hol} A. S. Holevo, {\em Statistical Structure of
Quantum Theory}, (Springer-Verlag, Berlin, 2001).
\bibitem{bra} V. B. Braginsky, Yu. I. Vorontsov, K. S. Thorne,
Science 209,
547 (1980); M. F. Bocko and R. Onofrio, Rev. Mod. Phys. 68, 755
(1996). See also the special issue on QND, Appl. Phys. B 64,
123
(1997), J. Mlynek et al. (Eds.).
\bibitem{dec04} T. C. Ralph, S. D. Bartlett, J. L. O'Brien, G.
J. Pryde, and
H. M. Wiseman, 
Phys. Rev. A {\bf 73}, 012113 (2006)
\bibitem{qb1} P. Kok, H. Lee, and J. P. Dowling, Phys. Rev. A {\bf 66}, 
063814 (2002). 
\bibitem{qb2} G. J. Pryde, J. L. O'Brien, 
A. G. White, S. D. Bartlett, and T. C. Ralph,
Phys. Rev. Lett. {\bf 92}, 190402 (2004). 
\bibitem{ulr} U. L. Andersen, M. Sabuncu, R. Filip, and G. Leuchs,
Phys. Rev. Lett. {\bf 96}, 020409 (2006)
\bibitem{mst} L. Mi$\check{\mbox{s}}$ta quant-ph/0510191.
\bibitem{filip} L. Mi$\check{\mbox{s}}$ta, R. Filip,
Phys. Rev. A {\bf 72}, 034307 (2005).
\bibitem{nap} A. Ferraro, S. Olivares, M. G. A. Paris, {\em
Gaussian states in Quantum Information}, (Bibliopolis, Napoli, 2005).
\bibitem{CVOperations} S.D. Bartlett, B.C. Sanders,S.L.
Braunstein, K. Nemoto, Phys. Rev. Lett. {\bf 88}, 097904 (2002).
\bibitem{msa} G. M. D'Ariano, M. F. Sacchi, Phys. Lett. A {\bf 231}, 325 (1997).
\bibitem{xqn}Matteo G. A. Paris, Phys. Rev. A, {\bf 65}, 012110 (2002).
\bibitem{ls1}S.F. Pereira, et al., Phys. Rev. Lett. {\bf 72}, 214 (1994); 
K. Bencheikh, et al., Phys. Rev. Lett. {\bf 75}, 3422 (1995);
K. Bencheikh, et al., Phys. Rev. Lett. {\bf 78}, 34 (1997); 
R. Bruckmeier, et al., Phys. Rev. Lett. {\bf 78}, 1243 (1997); 
R. Filip, preprint quant-ph/0404010.
\bibitem{ls2} L.-M. Duan, et al., Phys. Rev. Lett. {\bf 85},
5643 (2000); A. Kuzmich and E.S. Polzik, Phys. Rev. Lett. {\bf 85}, 5639 (2000); B.
Julsgaard, et al., Nature {\bf 413}, 400 (2001); 
C. Schori, et al., Phys. Rev. Lett. {\bf 89}, 057903 (2002).
\end{thebibliography}
\end{document}